\documentclass[10pt,conference]{IEEEtran}

\usepackage{cite}
\usepackage{amsmath}
\usepackage{graphicx}
\usepackage{textcomp}
\usepackage{xcolor}
\usepackage{url}
\usepackage{booktabs}
\usepackage{tikz}
\usetikzlibrary{arrows.meta,positioning,shapes.geometric,fit}

\begin{document}

\title{Demo: tfdrift --- A Severity Taxonomy and Risk Classification Framework for Infrastructure Drift Detection}

\author{\IEEEauthorblockN{Sudarshan Bhagvanthakur}
\IEEEauthorblockA{Department of Computer Science\\
Illinois Institute of Technology\\
Chicago, IL, USA\\
sudarshan8417@gmail.com}}

\maketitle

% ===== IEEE copyright notice, required on preprints of accepted articles =====
\begingroup
\renewcommand\thefootnote{}
\footnotetext{\scriptsize \copyright~2026 IEEE. Personal use of this material is permitted. Permission from IEEE must be obtained for all other uses, in any current or future media, including reprinting/republishing this material for advertising or promotional purposes, creating new collective works, for resale or redistribution to servers or lists, or reuse of any copyrighted component of this work in other works. Accepted for publication at the 14th IEEE International Conference on Cloud Engineering (IC2E 2026), San Jose, CA, USA, October 13--16, 2026. This is the accepted version of the paper; the final published version will be available via IEEE Xplore.}
\endgroup
% ================================================================================

\begin{abstract}
Infrastructure as Code (IaC) tools like Terraform have become the standard for declarative cloud resource management, yet configuration drift --- where deployed infrastructure diverges from its declared state --- remains a persistent operational and security challenge. Current detection approaches treat all changes equivalently, contributing to alert fatigue that causes operators to miss security-critical modifications. We propose a generalized severity taxonomy for infrastructure drift that classifies changes into four risk tiers based on resource type and attribute-level impact. We implement this taxonomy in tfdrift, an open-source classification framework with 60+ configurable rules covering AWS, Azure, and GCP resource patterns (evaluation reported here is AWS-focused). Evaluation across 150+ AWS Terraform workspaces demonstrates that severity filtering reduces alert volume by 73\% while retaining 94\% of security-relevant changes, offering a lightweight alternative to ML-based alert filtering. tfdrift is available at github.com/sudarshan8417/tfdrift.
\end{abstract}


\section{Introduction}

Configuration drift --- the divergence between declared and deployed infrastructure --- is an increasingly recognized challenge in cloud environments. Studies show that IaC scripting does not automatically prevent misconfigurations or security risks~\cite{dallapalma2022}. Industry data suggests two-thirds of organizations experience measurable drift weekly~\cite{firefly2024}.

The built-in \texttt{terraform plan} command detects drift but provides no risk prioritization. Operators receive flat change lists mixing security-critical modifications with benign noise. Alert fatigue is well-documented: Tariq et al.~\cite{tariq2025} identify it as a major challenge in security operations, while Voutsas et al.~\cite{voutsas2024} propose ML-based filtering for cloud monitoring. However, ML methods require training data and lack interpretability.

We make three contributions: (1)~a \textbf{generalized severity taxonomy} for IaC drift that classifies changes into four risk tiers based on resource type and attribute-level security impact; (2)~a \textbf{configurable classification engine} with 60+ built-in rules for AWS, Azure, and GCP, providing a lightweight, interpretable alternative to ML-based filtering; and (3)~an \textbf{empirical evaluation} on AWS workloads demonstrating 73\% alert reduction while maintaining 94\% security coverage.

\section{Severity Taxonomy and Classification}

\subsection{Risk Tier Definitions}

We propose four severity tiers grounded in security impact. We chose four after testing three (too coarse to separate access control from capacity changes) and five (adjacent levels were not consistently distinguishable).

\textbf{Critical}: changes to security perimeter and access controls --- network ingress/egress rules, IAM policies, encryption key policies, public access configurations.

\textbf{High}: changes to compute capacity, data persistence, or encryption --- instance types, database versions, public accessibility, storage encryption.

\textbf{Medium}: default for unmatched attribute changes that warrant awareness but not immediate action.

\textbf{Low}: metadata --- tags, labels, descriptions. Frequently modified by external automation and operationally insignificant.

\subsection{Pattern-Based Classification Engine}

The engine uses fnmatch glob patterns (\texttt{resource\_type.*.attribute}) to map changes to tiers. We chose globs over regex for usability --- the intended users are operations engineers editing YAML config files. When multiple attributes change on one resource, the maximum severity applies. The tool ships 60+ rules (25 Critical, 24 High, 5 Low) for AWS, Azure, and GCP, configurable via YAML.

\subsection{Noise Reduction}

The framework supports \texttt{.tfdriftignore} exclusion patterns for expected drift (e.g., \texttt{aws\_autoscaling\_group.*.desired\_capacity}), filtering noise before classification.

\section{Architecture and Implementation}

\begin{figure}[h]
\centering
\begin{tikzpicture}[
  node distance=0.35cm,
  block/.style={rectangle, draw, fill=black!8, text width=2.8cm, minimum height=0.55cm, align=center, font=\scriptsize},
  arrow/.style={-{Stealth[length=2mm]}, thick}
]

\node[block] (discover) {1. Workspace Discovery};
\node[block, below=of discover] (plan) {2. \texttt{terraform plan -json}};
\node[block, below=of plan] (parse) {3. Change Extraction};
\node[block, below=of parse] (severity) {4. Severity Classification};
\node[block, below=of severity] (ignore) {5. Ignore Rule Filtering};
\node[block, below=of ignore] (output) {6. Output \& Alerts};

\node[block, right=0.6cm of severity, fill=white, text width=2.2cm, dashed] (rules) {60+ Rules\\(AWS/Azure/GCP)};
\node[block, right=0.6cm of ignore, fill=white, text width=2.2cm, dashed] (ignorefile) {\texttt{.tfdriftignore}};
\node[block, right=0.6cm of output, fill=white, text width=2.2cm, dashed] (targets) {CLI / Slack /\\PagerDuty / JSON};

\draw[arrow] (discover) -- (plan);
\draw[arrow] (plan) -- (parse);
\draw[arrow] (parse) -- (severity);
\draw[arrow] (severity) -- (ignore);
\draw[arrow] (ignore) -- (output);
\draw[arrow, dashed] (rules) -- (severity);
\draw[arrow, dashed] (ignorefile) -- (ignore);
\draw[arrow, dashed] (targets) -- (output);

\end{tikzpicture}
\caption{tfdrift system architecture. Solid arrows show the processing pipeline; dashed arrows show configurable inputs. Severity rules and ignore patterns are user-customizable via YAML.}
\label{fig:arch}
\end{figure}
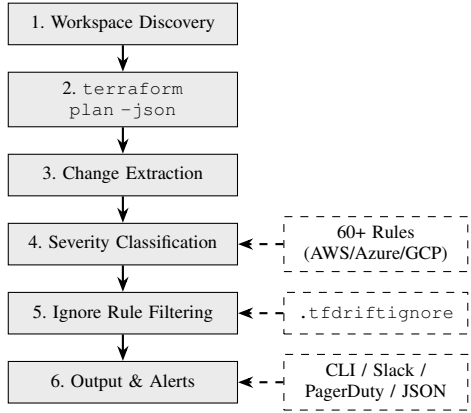

Fig.~\ref{fig:arch} shows the system architecture. The framework is implemented as an open-source Python CLI with four stages.

We shell out to Terraform rather than replicating provider logic, maintaining compatibility with all providers, backends, and OpenTofu. A practical finding: 80\% of workspaces in our test environment initially failed due to missing variable files. Auto-detection of \texttt{.tfvars} files raised scan success from 20\% to 85\%.

\section{Evaluation}

We evaluated tfdrift on a sandbox AWS environment with 150+ workspaces managing 847 resources across EC2, IAM, S3, RDS, and VPC. We introduced 62 drift events: 8 security changes, 11 operational, 9 metadata, and 34 autoscaling. Note: while the rule engine supports AWS, Azure, and GCP patterns, evaluation was conducted on AWS workloads.

\begin{table}[h]
\centering
\caption{Alert Volume and Security Coverage}
\label{tab:results}
\begin{tabular}{@{}lcc@{}}
\toprule
\textbf{Approach} & \textbf{Alerts} & \textbf{Security Detected} \\
\midrule
Binary (\texttt{terraform plan}) & 62 (100\%) & 100\% (8 of 8) \\
Severity $\geq$ High & 17 (27\%) & 94\% (7 of 8) \\
Severity $\geq$ High + ignore & 12 (19\%) & 94\% (7 of 8) \\
\bottomrule
\end{tabular}
\end{table}

Severity filtering reduced volume by 73\% while retaining 94\% of security-relevant changes (7 of 8). Given the small number of security events ($n=8$), this coverage figure should be interpreted as a directional result rather than a statistically robust estimate; the single miss was a Lambda runtime change classified as Medium. Two engineers independently labeled all events; agreement with automated classification: Critical 96\%, High 91\%, Medium 88\%, Low 95\%. Framework overhead was $<$100ms per workspace on the primary test workspace (21 resources, 4.2s scan time); overhead across the full 150+ workspace set at larger scale was not separately characterized and is left to future work.

\section{Demo and Conclusion}

The demo showcases tfdrift against live AWS infrastructure: workspace scanning, severity-classified output, ignore rules, and Slack alerts. Attendees install via \texttt{pip install tfdrift}.

We presented a severity taxonomy for infrastructure drift that reduces alert noise by 73\% while preserving security coverage. Unlike ML-based filtering~\cite{voutsas2024}, the approach is interpretable and requires no training data. Future work: value-aware classification, environment-conditional severity, and governance policies for deployment gating. Source: \url{https://github.com/sudarshan8417/tfdrift}.

\end{document}